\documentclass[smallextended]{svjour3}
\smartqed  
\usepackage{mathptmx,slashed}

\newcommand{\fant}[1]{\phantom{#1}}
\newcommand{\be}{\begin{equation}}
\newcommand{\ee}{\end{equation}}
\newcommand{\wdg}{\wedge}
\newcommand{\ot}{\otimes}

\journalname{GRG}

\begin{document}

\title{Quadratic Curvature  Gravity with Second Order Trace
and Massive Gravity Models in Three Dimensions}


\titlerunning{Quadratic Curvature  Gravity}        

\author{Ahmet Baykal}

\institute{A. Baykal\at
              Department of Physics, Faculty of Science and Letters, Ni\u gde University,  51240, Ni\u gde, Turkey\\
              \email{abaykal@nigde.edu.tr}           
}

\date{Received: date / Accepted: date}

\maketitle

\begin{abstract}
The quadratic curvature lagrangians having metric field equations with  second order trace  are constructed relative to an orthonormal coframe.
In $n>4$ dimensions, pure quadratic curvature lagrangian having second order trace  constructed contains
three free parameters in the most general case.  The fourth order field equations of some of these
 models, in arbitrary dimensions, are cast in a particular form using the Schouten tensor.  As a consequence,
the field equations for the New massive  gravity theory are related to those of the Topologically massive gravity.
In particular, the conditions under which the latter is ''square root" of the former are presented.

\end{abstract}

\section{Introduction}

Gravitational theories based on quadratic curvature (QC) invariants  were first considered long ago in the context
related to gauge invariance in \cite{weyl} as well as in \cite{eddington} soon  after the advent of general theory of relativity \cite{hj-scmidt-history}.
Since then their study have been motivated by various aspects of general theory of relativity and cosmology.
QC terms are also motivated by the effective action for the low energy limit of string theories \cite{zweibach}.
Although the QC terms lead to violation of unitarity, and ghost degree of freedom, they are favorable in the renormalization in the perturbative quantum gravity at one loop level \cite{odintsov-renor}. The gravitational models with QC terms also have desirable cosmological features such as they provide a framework for inflation \cite{starobinski} and with nonlinear curvature models it is possible to avoid cosmological singularity \cite{kerner}.
Nonlinear gravitational lagrangians also have the peculiar property that they can be cast into Einstein theory interacting with additional matter fields \cite{jakubiec}. Moreover, in contrast to the  cosmological models which introduces dark energy and dark matter, various nonlinear gravitational models provide alternatives to account for the observed late time acceleration of the Universe \cite{odintsov3}.

Currently, there has been a further interest in the modified gravitational theories that based on the actions involving   QC terms especially
in three dimensions. They are motivated by  the search for a consistent  model of quantum theory of gravity.
In particular, in three dimensions New massive gravity (NMG) \cite{NMG-bth} theory involves a particular combination of QC terms
complementing the usual Einstein-Hilbert action. In contrast to the topologically massive gravity (TMG) \cite{djt} which introduces mass  by addition of a Chern-Simons term to Einstein-Hilbert action, at the linearized level, NMG lagrangian is equivalent to Pauli-Fierz free, massive, spin-2 lagrangian.   In addition,   NMG theory also has parity-preserving field equations and it leads to a ghost-free quantum theory which is unitary at the  tree level. More recently, critical gravity models, that can be considered extension of NMG theory to four dimensions, lead to renormalizable gravity theory  by inclusion of  QC terms and a cosmological constant \cite{lu-pope}.

The NMG action has peculiar properties, such as the trace of the metric field equations yields back the QC lagrangian density and therefore the trace leads to second order partial derivatives in the metric components relative to a natural coframe. This particular property is crucial in showing that the linearized
equations are equivalent to  the Pauli-Fierz massive  lagrangian and hence introduction of mass to graviton. Recently, there are research  on extension of this property to QC lagrangians in higher dimensions \cite{nakasone-oda,ray}. The gravitational lagrangians having field equations of sixth order  partial derivatives of the metric components are investigated in order to single out lagrangian densities that are composed of cubic curvature terms
as well as terms having  square of the derivatives of the curvature tensor \cite{ray}. In \cite{nakasone-oda,ray} it has been  reported that in four dimensions, it is possible to construct QC actions whose metric field equations are of second order and the trace of the metric field equation  yielding back the quadratic part of the original action one started with. QC lagrangians having this particular property has been constructed explicitly by using three
curvature-squared terms, namely scalar curvature squared, Ricci-squared and Gauss-Bonnet term.
The resulting QC lagrangians contains \emph{two} free-parameters. Though, the generalization of NMG lagrangian to higher dimensions will be presented in arbitrary dimensions $n\geq3$ and  the lagrangian densities whose equations of motion  lead to second order trace are constructed using a different approach allowing spacetime dimensions $n=3$, $n=4$ and $n>4$ to be studied separately with the emphasis on the common structure of the fourth order lagrange multiplier terms. In particular, in three dimensions, the orthonormal coframe formulation of field equations allows one  to provide an interesting mathematical connection between  the field equations of the TMG and NMG theories.

The gravitational lagrangians  considered in what follows are based on QC invariants, in fact, represents a narrow sub-class of modified  models which include functions of these QC invariants. In particular, for the modified QC lagrangian of the form  $f(R, G)$ where  $G$ is the Gauss-Bonnet scalar (defined below as $G*1=\mathcal{L}^{(n)}_{GB}$ in $n$ dimensions) see the reviews, for example \cite{odintsov1} and \cite{odintsov2}.
In the following, it will be shown that, in general, such a modified QC model can be put into a form which is dynamically equivalent to original QC lagrangian
 coupled with additional scalar  field(s).

The outline of the paper is as follows.
In the next section,  the first order constrained formalism  is highlighted using the exterior forms relative to orthonormal coframe.
This section also serves to fix the notation that will be used throughout. The exterior differential form notation  and the geometrical conventions used  are taken from \cite{thirring,straumann}, see also the appendix in \cite{baykal-delice} for a concise introduction. The general formula presented are specialized to QC gravity models in the following two sections.
In the subsequent sections, the gravitational lagrangians in dimensions $n>4$ and  $n\leq4$ having  second order trace are constructed with emphasis on structure of their Lagrange multiplier term which contribute to the metric (or equivalently coframe) equations. $n>4$, $n=4$  and $n=3$ cases are discussed in the respective separate sections. In the section devoted to three dimensions, a formal mathematical relation between field equations of NMG and TMG  are  constructed in terms of the geometrical quantities. In particular, the conditions under which TMG is exact ``square root" of the NMG is presented.

\section{Variational derivatives relative  to an orthonormal coframe}
\label{sec:feqns}
Before delving into the explicit mathematical expression, for convenience, basic gravitational variables are defined briefly  here.
The basis coframe 1 -forms are denoted by $\theta^\alpha$ and the metric  tensor takes the form
$g=\eta_{\alpha\beta}\theta^\alpha\ot\theta^\beta$. The exterior product of the basis 1-forms will be  abbreviated as
$\theta^{\alpha}\wdg \theta^\beta\wdg\cdots\equiv\theta^{\alpha\beta\cdots}$. The covariant exterior derivative $D$ acts on tensor-valued forms.
The torsion 2-form $\Theta^\alpha=\frac{1}{2}T^{\alpha}_{\fant{a}\beta\mu}\theta^{\beta\mu}$ can be  defined as
\be
\Theta^\alpha=D\theta^\alpha=d\theta^\alpha+\omega^{\alpha}_{\fant{a}\beta}\wdg\theta^\beta
\ee
where $T^{\alpha}_{\fant{a}\beta\mu}$ are the components of the torsion tensor and $\omega^{\alpha}_{\fant{a}\beta}$ are connection 1-forms. Relative to an orthonormal coframe metric compatibility for the connection 1-forms reads $\omega_{\alpha\beta}+\omega_{\beta\alpha}=0$. In terms of Curvature 2-forms the Cartan's second structure equations reads
\be
\Omega^{\alpha}_{\fant{a}\beta}
=
\frac{1}{2}R^{\alpha}_{\fant{a}\beta\mu\nu}\theta^{\mu\nu}
=
d\omega^{\alpha}_{\fant{a}\beta}+\omega^{\alpha}_{\fant{a}\mu}\wdg \omega^{\mu}_{\fant{a}\beta}
\ee
where $R^{\alpha}_{\fant{a}\beta\mu\nu}$ are the components of the Riemann tensor.
The contractions of the forms and tensor-valued forms are defined with the help of the contraction operator
$i_{e_\alpha}\equiv i_\alpha$ where $e_\alpha$ is basis frame  fields that are
metric dual to the basis 1-forms: $i_\alpha\theta^\beta=\delta^{\beta}_{\alpha}$. $*$ is Hodge dual  that defines an inner product for two $p$-forms
in terms of the metric tensor. In terms of the Hodge dual, the oriented volume element reads $*1=\sqrt{|g|}dx^{0}\wdg\cdots\wdg dx^{(n-1)}$ in $n$ dimensions.
Ricci 1-form can be defined as $R^{\alpha}\equiv i_\beta\Omega^{\beta\alpha}=R^{\beta\alpha}_{\fant{aa}\beta\mu}\theta^\mu$
whereas the scalar curvature $R$ can be written as $R=i_\alpha R^\alpha$. All the other tensorial objects will be introduced as they required in terms of the
basic quantities defined here.

 In this section the  scheme  of calculation of variational derivative relative to an orthonormal coframe will briefly presented \cite{baykal-delice,var-dereli-tucker}. In later sections,  the general formulae of this section will be applied to all considerations in what follows.
As will be apparent in the following sections, the formulation of the QC metric field equations relative to an orthonormal coframe
leads to compact form that has an advantage  in the algebraic manipulations involved compared to the similar calculations relative to a coordinate basis, cf. for example, \cite{tekin-deser}.

The  gravitational lagrangian densities that will be studied below are all of the form
\be\label{lag-def}
\mathcal{L}^{(n)}
=
\mathcal{L}^{(n)}[\theta^\alpha, \omega^{\alpha}_{\fant{a}\beta}]
\ee
are $n$-forms  in $n$ dimensional Lorentzian manifold. The first order constrained formalism where coframe and connection 1-forms are
taken as independent gravitational variables \cite{kopczynsky,hehl} will be used.
All the lagrangians considered will therefore depend on the basic gravitational variables,  coframe and connection 1-forms. The lagrangians will depend on exterior derivative  of these variables, namely,  $d\theta^\alpha $ and $d\omega^{\alpha}_{\fant{a}\beta}$ through the torsion and curvature forms respectively.

The corresponding field equations  for the Riemannian metric is deduced from the coframe
variation of the lagrangian (\ref{lag-def}), subject to the condition that the torsion tensor $\Theta^\alpha=D\theta^\alpha$ vanishes.
Relative to an orthonormal coframe where the metric components are constants, the metric compatibility of the connection 1-forms reads
$\omega_{\alpha\beta}+\omega_{\beta\alpha}=0$. This algebraic condition can simply be imposed by antisymmetrizing
the coefficients of $\delta\omega_{\alpha\beta}$ whereas the torsion-free condition, $\Theta^\alpha=0$ which is a dynamical constraint on the gravitational variable $\theta^\alpha$,   can be implemented into the resultant field equations by extending the original lagrangian density to include a Lagrange multipliers $(n-2)$-forms $\lambda^\alpha$ as
\be
\mathcal{L'}^{(n)}[\theta^\alpha, \omega^{\alpha}_{\fant{a}\beta},\lambda_\alpha]
=
\mathcal{L}^{(n)}
+
\lambda^\alpha\wdg \Theta_\alpha.
\ee
The total variational derivative of the extended lagrangian density with respect to independent variables has, up to a disregarded boundary term,
the general form
\begin{eqnarray}
\delta\mathcal{L'}^{(n)}[\theta^\alpha, \omega^{\alpha}_{\fant{a}\beta},\lambda_\alpha]
&=&
\delta\omega_{\alpha\beta}\wdg[\Pi^{\alpha\beta}-\frac{1}{2}(\theta^\alpha\wdg \lambda^\beta-\theta^\beta\wdg \lambda^\alpha)]
\nonumber\\
&&+
\delta\theta^{\alpha}\wdg (*\mathcal{G}_\alpha+D\lambda_\alpha)
+
\delta\lambda^\alpha\wdg \Theta_\alpha\label{var-der-general}
\end{eqnarray}
where the convenient definitions
\be\label{generalized-var-derivative-expressions}
\Pi^{\alpha\beta}\equiv\frac{\delta\mathcal{L}^{(n)}}{\delta\omega_{\alpha\beta}}, \qquad
*\mathcal{G}^\alpha\equiv\frac{\delta\mathcal{L}^{(n)}}{\delta\theta_{\alpha}}
\ee
have been introduced. It is also convenient to define $(n-1)-$form $*E^\alpha\equiv\delta\mathcal{L'}^{(n)}/\delta\theta^\alpha \linebreak=*\mathcal{G}^\alpha+D\lambda^\alpha$.
It is symmetric as a result of the local Lorentz invariance of the basis coframe: $\theta^\beta\wdg *E^\alpha-\theta^\alpha\wdg *E^\beta=0$ and it is covariantly constant as a result of diffeomorphism invariance of  lagrangian composed of curvature scalars: $D*E^\alpha=0$.

In the orthonormal coframe formulation, the metric field equations derived from coframe variational derivative, namely the vacuum equations $*E^\alpha=0$ are defined by (\ref{var-der-general}). The vacuum field equations $*E^\alpha=0$ involve the exterior covariant derivative of the lagrange multiplier term which can be expressed in terms of the other independent fields by making use of the variational derivatives of the extended  lagrangian with respect to the connection 1-forms. The lagrange multiplier term, $\lambda^\alpha$ is a vector-valued $(n-2)$-form while $\Pi^{\alpha\beta}$ is antisymmetric bivector-valued $(n-1)$-form and they  have equal number of components \cite{hehl}. The   equivalence can be expressed  precisely as
\be\label{general-expression-lag-mult}
\lambda^\beta
=
2i_\alpha\Pi^{\alpha\beta}+\frac{1}{2}\theta^\beta\wdg i_\mu i_\nu \Pi^{\mu\nu}
\ee
which can be obtained by calculating two successive contractions of  equations derived from variational derivative of
 the extended lagrangian  with respect to  the connection  subject to the constraint vanishing torsion constraint.  $\Theta^\alpha=0$ results from the variational derivative of the extended lagrangian with respect to the lagrange multiplier $(n-2)$-forms.
Because the  contribution of the Lagrange multiplier  term to the metric field equations induced by the coframe variations is given by $D\lambda^\alpha$, the properties  of the tensor-valued $(n-1)$ form $\Pi^{\alpha\beta}$ becomes crucial for the study of lagrangian densities below since it determines  the fourth order terms. Hence, it is convenient to  derive some of the general properties that will be used later.

The contribution of the Lagrange multiplier  to the trace of metric field equations is $\theta^\alpha\wdg D\lambda_\alpha$ which can also be expressible in terms of the bivector valued form $\Pi^{\alpha\beta}$ using the general expression (\ref{general-expression-lag-mult}) for it as
\be\label{gen-trace-formula}
 \theta_\alpha\wdg D\lambda^\alpha
 =
-2Di_\beta(\theta_\alpha \wdg  \Pi^{\beta\alpha})
\ee
 In particular, this important  result implies that in the case in which $\Pi^{\beta\alpha}$ satisfy $\theta_\beta \wdg  \Pi^{\beta\alpha}=0$, the Lagrange multiplier terms do not contribute to the trace of the metric field equations.

 As another property, the invariance under the general coordinate transformations require $D*E^\alpha=0$ which involves
 $D^2\lambda^\beta$. This term  can conveniently be expressed in terms of bivector-valued form $\Pi^{\alpha\beta}$ as
\be\label{D2-lambda}
D^2\lambda^\beta
=
2\Omega^\beta_{\fant{a}\alpha}\wdg i_\mu\Pi^{\mu\alpha}
\ee
where the first Bianchi identity, $D^2\theta^\beta=\Omega^\beta_{\fant{a}\alpha}\wdg\theta^\alpha=0$, has been used.

The covariant exterior derivative of bivector-valued 1-form can be calculated using the field equations for the connection 1-form. In a metric theory,
$D$ corresponds to covariant exterior derivative  of torsion-free metric compatible connection  1-form and  the covariant exterior derivative of the equations for the connection 1-form yields
\be\label{cov-der-pi}
D\Pi^{\alpha\beta}
=
-\frac{1}{2}(\theta^{\alpha}\wdg D\lambda^{\beta}-\theta^{\beta}\wdg D\lambda^{\alpha})
\ee
after $D\theta^\alpha=\Theta^\alpha=0$ has been used.
It is interesting to note that the right hand side of (\ref{cov-der-pi}) vanishes identically for the QC lagrangians having second order trace studied below.
The general scheme of calculations presented here is specialized  to QC lagrangians in the following sections.

\section{Field equations for QC models and their trace}

As will be apparent in what follows, the field equations for QC lagrangians depend on the dimensionality of the spacetime $n$ and therefore
the dimensions $n>4$ and $n=4$ and $n=3$ are to be studied separately.
The presentation of the field equations for the QC lagrangians allows one to study the field equations in its most general form.
It is convenient to start with $n>4$ dimensions and then specialize to lower dimensions.
In $n>4$ spacetime dimensions, the most general QC gravitational lagrangian will be parameterized by the following individual QC lagrangians
\begin{eqnarray}
\mathcal{L}^{(n)}_{t}
&=&
a\mathcal{L}^{(n)}_{1}
+
b\mathcal{L}^{(n)}_{2}
+
c
\mathcal{L}^{(n)}_{3}
\nonumber
\\
&=&
aR^2*1+bR^\alpha\wdg*R_\alpha+c\Omega_{\alpha\beta}\wdg*\Omega^{\alpha\beta}
\label{n>4-qc-lag}
\end{eqnarray}
involving the linear combinations  all three of the QC scalars with three coupling parameters $a, b, c$. (The numerals and the Latin subscripts for lagrangian densities are not tensorial and are used as discriminating labels.)
Using the field equations given in the previous section, the equations for the QC lagrangian  (\ref{n>4-qc-lag})  can be written in the form
\be\label{4+d-field-eqn-form1}
\Omega_{\alpha\beta}\wdg i^\mu*X^{\alpha\beta}_t-i^\mu\mathcal{L}^{(n)}_t+D\lambda_t^\mu=0
\ee
where $D\lambda_t^\mu=d\lambda_t^\mu+\omega^{\mu}_{\fant{a}\nu}\wdg \lambda_t^\nu$ and
the bivector-valued 2-form $X_t^{\alpha\beta}$ is given by
\be\label{aux-X-n+d}
X_t^{\alpha\beta}
\equiv
2c\Omega^{\alpha\beta}
+
\theta^{\alpha}\wdg (bR^\beta-aR\theta^{\beta})-\theta^\beta\wdg (bR^\alpha-aR\theta^{\alpha}).
\ee
Since the field equations (\ref{4+d-field-eqn-form1}) do not contain the number of dimensions $n$ explicitly, they are, in fact, valid  for all $n$.
For the calculations in the following sections, the auxiliary anti-symmetric tensor-valued 2-form $X^{\alpha\beta}$ is very convenient and it is also
to be used in the calculation of lagrange multiplier term in terms of the gravitational field variables as well.
For QC lagrangians, it has the general expression as the linear  combinations of curvature 2-forms and Ricci 1-forms and the scalar curvature.
Explicitly, it can equivalently be introduced by the defining relation
\be
\Pi^{\alpha\beta}
\equiv
D*X^{\alpha\beta}
=
d*X^{\alpha\beta}
+
\omega^{\alpha}_{\fant{a}\mu}\wdg *X^{\mu\beta}
+
\omega^{\beta}_{\fant{a}\mu}\wdg *X^{\alpha\mu}.
\ee
 The general QC metric field equations (\ref{4+d-field-eqn-form1})  then can be cast  into the following remarkably concise  form
\be\label{4+d-field-eqn}
D\lambda_t^\mu
+
*T^\mu_t[R, R^\alpha,\Omega^{\alpha\beta}]
=0
\ee
which separates naturally the terms that are linear in curvature (the lagrange multiplier terms) and QC terms.
In terms of the contraction of the auxiliary 2-form $X_t^{\alpha\beta}$ and with the curvature 2-form,
 $(n-1)$-form  $*T^\mu_t$ can be defined as
\be\label{qc-en-mom}
*T^\mu_t[R, R^\alpha,\Omega^{\alpha\beta}]
\equiv
-\frac{1}{2}(i^\mu\Omega_{\alpha\beta})\wdg *X^{\alpha\beta}_t+\frac{1}{2}\Omega_{\alpha\beta}\wdg i^\mu*X^{\alpha\beta}_t
\ee
for the total QC lagrangian in (\ref{n>4-qc-lag}). The auxiliary  $(n-1)$-form $*T^\mu_t$  has  a mathematical structure that is formally analogous  to energy-momentum form of  a generic minimally coupled matter 2-form field, $F=\frac{1}{2}F_{\alpha\beta}\theta^{\alpha\beta}$, expressed relative to an orthonormal coframe (see, e.g., \cite{thirring,straumann}) and henceforth will be called
energy-momentum $(n-1)$-form for the QC terms.

In the general case corresponding to (\ref{n>4-qc-lag}), the trace of the energy-momentum form $*T^\mu_t[R, R^\alpha,\Omega^{\alpha\beta}]$ can be found as
\be
T_{t}
\equiv
T^\mu_{t\fant{.}\mu}[R, R^\alpha,\Omega^{\alpha\beta}]*1
=
\frac{1}{2}(n-4)\Omega^{\alpha\beta}\wdg *X_t^{\alpha\beta}.
\ee
The general trace expression vanishes identically for $n=4$ which in fact reflects the dimensional dependent nature of the  general QC lagrangian (\ref{n>4-qc-lag}). Moreover, note here that $\mathcal{L}^{(n)}_{t}$ can conveniently be rewritten  in terms of contractions of auxiliary 2-form $X^{\alpha\beta}_t$ with curvature 2-form as
\be\label{n+4lagrangian-second-form}
\mathcal{L}^{(n)}_{t}
=
\frac{1}{2}\Omega_{\alpha\beta}\wdg * X_t^{\alpha\beta}
\ee
also that the lagrange multiplier term in (\ref{4+d-field-eqn}) which is  linear in curvature and it comprises  all the  terms that are fourth order in the partial derivatives of the metric components relative to a coordinate basis. Therefore, the form of the field equation (\ref{4+d-field-eqn}) suggest that the trace of the field equations is    either fourth or second order in partial derivatives of the metric and the only way to have a second order trace is to have $D\lambda^\mu_t$ term to have vanishing trace. In the latter case, the trace becomes proportional to the general lagrangian density (\ref{n>4-qc-lag}) as will be elaborated below. Finally, note that the covariant exterior derivative (or equivalently the divergence) of the energy-momentum form can be calculated,  by making use of $\Pi_{t}^{\alpha\beta}=D*X_t^{\alpha\beta}$ in  the general formulae (\ref{D2-lambda}) of the previous section.

The addition of Einstein-Hilbert and/or cosmological terms is of considerable interest in applications and will be considered in  especially in the sections devoted to three dimensions and the expression of the energy-momentum form for QC terms in terms of the auxiliary 2-form $X^{\alpha\beta}$ then allows one to write down the  corresponding terms in the field equations. Explicitly, if an Einstein-Hilbert lagrangian density of the form $\frac{\sigma}{2}R*1$ is added to the general QC lagrangians (\ref{n>4-qc-lag}), then the corresponding field equations can effectively be obtained  by the formal replacement
\be
X^{\alpha\beta}_t
\mapsto
X^{\alpha\beta}_t+\frac{\sigma}{2}\theta^{\alpha\beta}
\ee
($\sigma$ and $\kappa$ are constants) in (\ref{4+d-field-eqn-form1}). It  this convenient to recall at this point that the Einstein $(n-1)$-form
$*G^\alpha=*(R^\alpha-\frac{1}{2}R\theta^\alpha)$ can be expressed in terms of  the contraction  $*G^\alpha=-\frac{1}{2}\Omega_{\alpha\beta}\wdg i^\mu*\theta^{\alpha\beta}$. Addition  of  $n$-form  $\lambda*1$ to (\ref{n>4-qc-lag}), with  $\lambda$ being a cosmological constant,  leads to the replacement
\be
*T^\mu_t\mapsto *T^\mu_t+\lambda*\theta^{\mu}
\ee
for the energy-momentum terms of the QC terms in the general field equations (\ref{4+d-field-eqn}).
These formal replacements apply to any QC lagrangian in $n\geq3$ dimensions studied below.

It is well known that the particular QC $n-$form, namely the Gauss-Bonnet  term
\be
\mathcal{L}^{(n)}_{GB}
=
\frac{1}{2}\Omega_{\alpha\beta}\wdg \Omega_{\mu\nu}\wdg *\theta^{\alpha\beta\mu\nu}
\ee
leads to second order field equations in the partial derivatives of the metric and consequently the corresponding trace of the metric equations is also second order \cite{lovelock}. In four dimensions, Gauss-Bonnet  the variational derivative vanishes identically whereas in three dimensions, the term itself vanishes identically.
 Although the Gauss-Bonnet QC term will not be considered below such a term can easily be included
into the general analysis in $n>4$ dimensions, for which the variational derivative with  respect to connection 1-form leads to
\be
\Pi^{\alpha\beta}_{GB}
=
D(\Omega_{\mu\nu}\wdg *\theta^{\alpha\beta\mu\nu}).
\ee
and this expression vanishes identically as a result of second Bianchi identities. In turn, the corresponding lagrange multiplier
$(n-1)$-form $\lambda^\mu_{GB}$ vanishes identically. Therefore the contribution  of $\mathcal{L}_{GB}$ to coframe variational derivative with respect to $\delta\theta_\lambda$ is
second order in the partial derivatives of the metric components and is of the form $\frac{1}{2}\Omega_{\mu\nu}\wdg\Omega_{\alpha\beta}\wdg*\theta^{\alpha\beta\mu\nu\lambda}.$

Next, it is convenient   to the study the trace of QC field equations in the most general case.
by wedging the metric field Eqn. (\ref{4+d-field-eqn}) from the left by basis coframe 1-form $\theta_\mu$ and using (\ref{n+4lagrangian-second-form}),  it is possible to show that the trace of the metric equations simplifies to the expression
\be
\frac{1}{2}(n-4)\Omega_{\alpha\beta}\wdg*X^{\alpha\beta}_t
+
\theta_{\mu}\wdg D\lambda_t^\mu=0.
\ee
The Lagrange multiplier term can further be simplified  by making use of the general formulae (\ref{gen-trace-formula}) with $\Pi^{\alpha\beta}$ specialized to that of derived from $X^{\alpha\beta}_t$ in (\ref{aux-X-n+d}) as
\be
\theta_{\mu}\wdg \lambda_t^\mu
=
2i_\alpha D*\left\{[(n-2)b+2c]R^\alpha+[2(n-1)a+b]R\theta^\alpha\right\}
\ee
where $\Pi^{\alpha\beta}_t+\Pi^{\beta\alpha}_t=0$ has been used.
Consequently, the trace of the general field equations for QC lagrangian takes the form
\be
(n-4)\mathcal{L}^{(n)}_t-2Di_\alpha D*\left\{[(n-2)b+2c]R^\alpha+[2(n-1)a+b]R\theta^\alpha\right\}
=0.
\ee
 As a result for all the QC gravitational actions, the number of spacetime dimensions $n=4$ turns out to be critical dimension. In particular, for $n=4$
 the trace equations is simply the homogeneous wave equation for the scalar curvature in the most general case. The contribution of the
 Lagrange multiplier form involves the covariant exterior derivatives of terms linear in Ricci forms and scalar curvature. As mentioned above, this implies that there are in general two different  cases regarding the trace in which it involves  fourth and second order partial derivatives of the metric.

In the fourth order case where the Lagrange multiplier term contributes to the trace,  the trace can further  be simplified by making use of the contracted second Bianchi identity $D*G^\alpha=0$. In doing so, the trace then takes the form
 \be\label{4th-order-trace}
(n-4)\mathcal{L}^{(n)}_t-[4(n-1)a+nb+2c]d*dR
=0
\ee
assuming  that the coefficient of the $d*dR$ term does not vanish identically.
Even in this case, the number of spacetime dimension $n=4$ is singled out  with  regard to the trace of the  field equations for the most general QC lagrangian.  For $n=4$, (\ref{4th-order-trace}) reduces to  $d*dR=0$, unless the coefficient in the square brackets is zero and becomes fourth order in the derivative of the metric components. The subcases for which the term in the square brackets vanishes identically will briefly be studied
from slightly different point of view below.

In the case where the trace is second order in the derivatives of the metric, the contribution of the Lagrange  multiplier term can be made to vanish by tuning the coupling parameters $a,b,c$.  Note also that from the result in (\ref{4th-order-trace}) that in this case the trace of the field equations yield back the  lagrangian from which the field equations are derived up to a numerical constant depending on the dimension. In $n>4$ spacetime dimensions, there are three distinct subcases each of which corresponds to QC lagrangians that are inequivalent.
These cases can be derived  as linear combinations of the QC lagrangians $\mathcal{L}_k$ as follows.

\begin{itemize}
\item[(1)]$c=0$, $\mathcal{L}^{(n)}_{12}\equiv a\mathcal{L}^{(n)}_{1}+b\mathcal{L}^{(n)}_{2}$
for which  the metric field equations read
\be
\Omega_{\alpha\beta}\wdg i^\mu*[\theta^{\alpha}\wdg(bR^\beta+aR\theta^\beta)-\theta^{\beta}\wdg(bR^\alpha+aR\theta^\alpha)]
-
i^\mu\mathcal{L}^{(n)}_{12}+D\lambda^\mu_{12}=0
\ee
with the corresponding auxiliary bivector-valued forms given by
\be
\Pi_{12}^{\alpha\beta}
=
D*[\theta^{\alpha}\wdg(bR^\beta+aR\theta^\beta)-\theta^{\beta}\wdg(bR^\alpha+aR\theta^\alpha)].
\ee
Consequently, using this auxiliary 2-form given  in (\ref{general-expression-lag-mult}), it is possible to show the expression for Lagrange multiplier form can be simplified to
\be\label{lag-mult12-form2}
\lambda_{12}^\alpha
=
*D\left[2bR^\alpha+\left(2a+\frac{b}{2}\right)R\theta^\alpha\right].
\ee
Using the general considerations of the previous section, the trace of the metric equations can be found as
\be\label{quad-trace-12}
(n-4)\mathcal{L}^{(n)}_{12}-2Di_\alpha D*\{(n-2)bR^\alpha+[2(n-1)a+b]R\theta^\alpha\}=0.
\ee
In contrast to the fourth order trace-case above, instead of using the contracted second Bianchi identity to simplify the trace expression, in order for the trace to be of second order, one requires that the lagrange multiplier term to be proportional to the contracted Bianchi identity.
This can only happen when  the expression inside the curly brackets above becomes proportional to Einstein 1-form.
In other words,  with the judicious choice of the constants
\be\label{quad12-a-b-cond}
\frac{a}{b}
=
-
\frac{n}{4(n-1)}
\ee
the trace expression reduces  to $(n-4)\mathcal{L}^{(n)}_{12}=0$ as a result of $D*G^\mu=0$. This subcase also turns out to be also
persist in  three and four dimensions as well as will be presented below and therefore the result generalizes in fact to dimensions
$n\geq3$.
It is worth noting that the QC lagrangian density $\mathcal{L}^{(n)}_{12}$,
 with the coupling constants satisfying (\ref{quad12-a-b-cond}),  (\ref{lag-mult12-form2}) yields
$\lambda^\alpha\propto*DL^\alpha$ and that it is possible to rewrite QC lagrangian $\mathcal{L}^{(n)}_{12}$ in the following equivalent forms
\be\label{schouten-squared-lagrangian-form2}
\mathcal{L}^{(n)}_{12}\left[a
=
-
\frac{n}{4(n-1)},b=1\right]
=
-L_\alpha\wdg L_\beta\wdg *\theta^{\alpha\beta}
=
-G^\alpha\wdg *L_\alpha
\ee
where $L^\alpha=R^\alpha-\frac{1}{2(n-1)}R\theta^\alpha$ is the Schouten 1-form, defined in $n\geq3$ dimensions \cite{deser} and
$G^\alpha=R^\alpha-\frac{1}{2}R\theta^\alpha$ is  the Einstein 1-form.
Various properties of  (\ref{schouten-squared-lagrangian-form2})  such as the variational derivatives with respect to $L^\alpha$ and $G^\alpha$ have been studied in \cite{heinicke} in more general  context of Riemann-Cartan geometry using the language of exterior differential forms.

Finally, note that,  with the addition of an appropriate  cosmological constant to  (\ref{schouten-squared-lagrangian-form2}), the lagrangian density becomes the critical theory \cite{lu-pope} which has recently  been introduced for $n=4$.

\item[(2)]$b=0$, $\mathcal{L}^{(n)}_t=\mathcal{L}^{(n)}_{13}=a\mathcal{L}^{(n)}_{1}+c\mathcal{L}^{(n)}_{3}$
has the metric equations
\be
2\Omega_{\alpha\beta}\wdg i^\mu*(c\Omega^{\alpha\beta}+aR\theta^{\alpha\beta})
-
i^\mu\mathcal{L}^{(n)}_{13}+D\lambda^\mu_{13}=0
\ee
with the corresponding auxiliary bivector-valued forms given by
\be
\Pi_{13}^{\alpha\beta}
=
2D*(c\Omega^{\alpha\beta}+aR\theta^{\alpha\beta}).
\ee
and thus the trace is now given by
\be
(n-4)\mathcal{L}^{(n)}_{13}-2Di_\alpha D*[cR^\alpha+(n-1)aR\theta^\alpha]=0.
\ee
Analogous to  the  first case, if the two non-zero constants  are related by
\be\label{quad13-a-b-cond}
\frac{a}{c}
=
-
\frac{1}{(n-1)}
\ee
then trace expression again reduces  to $\mathcal{L}^{(n)}_{13}=0$ by means  of the contracted second Bianchi identity $D*G^\mu=0$.

\item[(3)]$a=0$, $\mathcal{L}^{(n)}_t\equiv\mathcal{L}^{(n)}_{23}=b\mathcal{L}^{(n)}_{2}+c\mathcal{L}^{(n)}_{3}$.
The metric field equations that follow from this density takes the form
\be
\Omega_{\alpha\beta}\wdg i^\mu*[2c\Omega^{\alpha\beta}+b(\theta^{\alpha}\wdg R^\beta-\theta^{\beta}\wdg R^\alpha)]
-
i^\mu\mathcal{L}^{(n)}_{23}+D\lambda^\mu_{23}=0
\ee
with the corresponding auxiliary bivector-valued forms given by
\be
\Pi_{23}^{\alpha\beta}
=
D*[2c\Omega^{\alpha\beta}+b(\theta^{\alpha}\wdg R^\beta-\theta^{\beta}\wdg R^\alpha)].
\ee
As in the previous two cases, for the trace of the field equations to be of second order, the constants $b, c$ are to be chosen such that the Lagrange multiplier terms add up to contracted second Bianchi identity. Explicitly, the trace of the metric field equations are
\[
(n-4)\mathcal{L}^{(n)}_{23}-2Di_\alpha D*\{[2c+(n-2)b] R^\alpha +bR\theta^\alpha\}=0
\]
Thus, if the constants $b$ and $c$ satisfy  $2c+nb=0$, the fourth order Lagrange multiplier term vanishes identically.
\end{itemize}

\section{Modified QC models}
In order to study modified QC models which involve  some functions of  QC terms  relative to an orthonormal coframe, one first defines
the relevant QC scalars  $Q, P, K$ by making use of (\ref{n>4-qc-lag}) as
\be
\mathcal{L}^{(n)}_1\equiv Q*1,\qquad \mathcal{L}^{(n)}_2\equiv P*1, \qquad \mathcal{L}^{(n)}_3\equiv K*1.
\ee
Now, it is convenient to consider the  gravitational lagrangian that depends on the
quadratic curvature invariants $Q, P, K$  in the form
\be\label{mod-qc-lag-form2}
\mathcal{L}^{(n)}_{mod.}
=
f(Q, P, K)*1
\ee
with the assumption that $f$ satisfies some regularity condition stated below.
Note that although it is possible to start  with a function $f$ depends also on other QC terms, such as,  Gauss-Bonnet term
and/or Weyl curvature-squared term. For the sake of the simplicity of the argument, such terms will not be considered.

It is  easy to write the total variational derivative of the lagrangian $\mathcal{L}^{(n)}_{mod.}$
in the form
\be\label{mod-var-form1}
\delta\mathcal{L}^{(n)}_{mod.}
=
f_Q\delta\mathcal{L}^{(n)}_1+f_P\delta\mathcal{L}^{(n)}_2+f_K\delta\mathcal{L}^{(n)}_3+ \left[f-Qf_Q-Pf_P-Kf_K\right]\delta*1.
\ee
The second term on the right hand side  contributes to coframe equations only  and in particular it
is, by definition, the Legendre transform of the function $f(Q, P, K)$. This identification then allows one to introduce
the scalar fields
\be
\phi_1\equiv f_Q=\frac{\partial f}{\partial Q},\qquad\phi_2\equiv f_P=\frac{\partial f}{\partial P},\qquad\phi_3\equiv f_K=\frac{\partial f}{\partial K}.
\ee
Consequently,
the variational derivative of the modified lagrangian $\mathcal{L}^{(n)}_{mod.}$  in (\ref{mod-var-form1}) becomes equivalent to that of the Scalar-Tensor(ST) type lagrangian
\be
\mathcal{L}^{(n)}_{ST}
=
\phi_1\mathcal{L}^{(n)}_1+\phi_2\mathcal{L}^{(n)}_2+\phi_3\mathcal{L}^{(n)}_3+ V(\phi_1,\phi_2,\phi_3)*1
\ee
provided that the potential term $V(\phi_1,\phi_2,\phi_3)$ for the scalar fields is taken to be the Legendre transform of $f(Q, P, K)$ up to an irrelevant sign, namely
\be
V(\phi_1,\phi_2,\phi_3)
\equiv
f-Qf_Q-Pf_P-Kf_K.
\ee
Similar to the well-known $f(R)$ theory case, a dependence of the function $f$ on a particular QC invariant introduces a corresponding scalar field via an appropriate Legendre transform. The change of variables $\{f_Q, f_P, f_K\}\mapsto \{\phi_1, \phi_2, \phi_3\}$ requires that the Legendre   transform to be non-singular. The fulfilment of the requirement however depends on the explicit form of the function $f$.

In the above general framework, the dynamically equivalent ST form of the field equations  for the modified  gravitational lagrangian (\ref{mod-qc-lag-form2}) can concisely  be written as
\begin{eqnarray}
&\phi_1 *E^\alpha_1+\phi_2 *E^\alpha_2+\phi_3 *E^\alpha_3+V(\phi_1, \phi_2, \phi_3)*\theta^\alpha
\nonumber\\
&
+
2d\phi_1\wdg i_\beta D*X^{\alpha\beta}_1
+
2D[(i_\beta d\phi_1)*X^{\alpha\beta}_1]
\nonumber\\
&
+
2d\phi_2\wdg i_\beta D*X^{\alpha\beta}_2
+
2D[(i_\beta d\phi_2)*X^{\alpha\beta}_2]
\nonumber\\
&
+
2d\phi_3\wdg i_\beta D*X^{\alpha\beta}_3
+
2D[(i_\beta d\phi_3)*X^{\alpha\beta}_3]=0\label{general-modified-eqn}
\end{eqnarray}
where $*E^\alpha_k=0$ are the field equations corresponding to the QC lagrangian densities $\mathcal{L}^{(n)}_k$ for $k=1,2,3$ respectively
and $X^{\alpha\beta}_k$ are the corresponding bi-vector valued forms defined in the previous section.

The trace of the metric field equations (\ref{general-modified-eqn}) in this general case  now determines the equation that scalar fields satisfy.
However, the dynamical nature of the scalar fields depends crucially on the explicit form of the function $f(Q, P, K)$.

Finally, in the light of the above discussion, it is worth to point out that modification of the QC topological term $\mathcal{L}^{(4)}_{GB}=G*1$ to the form $\mathcal{L}^{(4)}_{mod. GB}=f(G)*1$ inevitably induces a scalar field and consequently, it is possible to write a dynamically  equivalent lagrangian density in the form $\phi \mathcal{L}^{(4)}_{GB}+V(\phi)*1$ in terms of the scalar field $\phi\equiv f_G$ defined as in the above examples.
The elaboration of ideas related to the ST equivalents  of such modified QC models  will be taken up elsewhere.

\section{$n>4$ dimensions}

Although the form of the field equations derived above are the same for all dimensions, apparently, the number of spacetime dimensions turns out to be   a  crucial parameter for the field equations and consequently in the calculation of their trace. Note that in $n> 4$ dimensions the QC lagrangian densities $\mathcal{L}^{(n)}_{12}$, $\mathcal{L}^{(n)}_{13}$ and  $\mathcal{L}^{(n)}_{23}$ are not equivalent since they can not be related by an addition of a total derivative QC term.

The results of the cases (1)-(3) allow one to construct  more general QC actions.
An arbitrary linear combination of the densities $\mathcal{L}^{(n)}_{12}$, $\mathcal{L}^{(n)}_{13}$ and $\mathcal{L}^{(n)}_{23}$ with \emph{three} arbitrary  constant constitutes the most general QC lagrangian density
with second order trace, or to put it more precisely, the trace is proportional the lagrangian density from which the field equations are derived (compare the construction, for example, given in \cite{ray}).  As a result of the above analysis of QC lagrangians, it is easy to construct the most general QC  lagrangian having second order trace in terms of $\mathcal{L}_{12}$, $\mathcal{L}_{13}$ and $\mathcal{L}_{23}$ with three free parameters. In the most general case, the total lagrangian density takes the form of the linear combination
\begin{eqnarray}\label{original-n-d-lag}
\mathcal{L}^{(n)}_g
&=&
p\mathcal{L}^{(n)}_{12}+q\mathcal{L}^{(n)}_{13}+r\mathcal{L}^{(n)}_{23}
\nonumber\\
&=&
-\frac{np+q}{4(n-1)}R^2*1+(p+r)R_\alpha\wdg*R^\alpha+(q-\frac{1}{2}nr)\Omega_{\alpha\beta}\wdg*\Omega^{\alpha\beta}
\end{eqnarray}
with $p, q, r$ being the free parameters of the pure QC lagrangian. Note that the field equations  for $\mathcal{L}^{(n)}_g$ are given by
\be\label{example}
\Omega_{\alpha\beta}\wdg i^\mu*
X_g^{\alpha\beta}
-
i^\mu\mathcal{L}^{(n)}_{g}+D\lambda^\mu_{g}=0
\ee
where
\begin{eqnarray}
X_g^{\alpha\beta}
=
(q-\frac{1}{2}nr)\Omega^{\alpha\beta}
&+&\theta^{\alpha}\wdg\left[(p+r)R^\beta-\frac{np+q}{4(n-1)}R\theta^\beta\right]
\nonumber\\
&-&
\theta^{\beta}\wdg \left[(p+r)R^\alpha-\frac{np+q}{4(n-1)}R\theta^\alpha\right]
\end{eqnarray}
and $\Pi_g^{\alpha\beta}
=
D*X_g^{\alpha\beta}$.
Note that the trace of the metric equations imply that when the metric field equations are satisfied the  action vanishes identically since the tracing the metric equations gives $\mathcal{L}^{(n)}_g=0$ for $n\neq4$. This result can in fact be used to further simplify the metric equations (\ref{example}).

In three and four dimensions, on the other hand,  only two of the cases out of total (1)-(3)  are independent and one of the cases can be written in terms of the remaining two.   Note also that, in all the cases, the existence of a second order trace, or equivalently, the vanishing of the lagrange multiplier terms, requires this term to involve the  Einstein form, in contrast, in three dimensions it  involves Schouten form. On the other hand, neither the case (2) nor (3) seems to have been investigated so far as the  QC lagrangian that generalizes the second order trace property of the NMG lagrangian to dimensions $n>4.$

As another important example of QC lagrangians in $n\geq 4$ dimensions, which will also be of interest in four dimension is the Weyl tensor-squared QC lagrangian density. From the point of view adopted above, it corresponds to the particular values of the parameters $p,q,r$.
Weyl 2-form $C^{\alpha\beta}=\frac{1}{2}C^{\alpha\beta}_{\fant{\alpha\beta}\mu\nu}\theta^{\mu\nu}$ can be defined in terms of curvature 2-forms and its contractions by \cite{thirring}
\be\label{weyl-def}
C^{\alpha\beta}
=
\Omega^{\alpha\beta}
-
\frac{1}{(n-2)}(\theta^{\alpha}\wdg R^\beta-\theta^{\beta}\wdg R^\alpha)+\frac{1}{(n-1)(n-2)}R\theta^{\alpha\beta}
\ee
for $n\geq 4$ dimensions. By contracting this equation side by side one finds that
\be\label{weyl-square-eqn-n-dim-form1}
\mathcal{L}^{(n)}_{W}
=
C_{\alpha\beta}\wdg*C^{\alpha\beta}
=
\Omega_{\alpha\beta}\wdg*\Omega^{\alpha\beta}-\frac{2}{(n-2)}R_\alpha\wdg *R^\alpha+\frac{1}{(n-1)(n-2)}R^2*1
\ee
which is suitable for the study of the field equations in the above framework.
The corresponding metric field equations can be written down immediately as
\be
D\lambda^\mu_W
+
*T^\mu_W
=0
\ee
with
$X^{\alpha\beta}_W=2*C^{\alpha\beta}.$
In this case, the lagrange multiplier term turns out to have a simple expression in terms of  Cotton  2-form as a result of the identity
\be
\Pi^{\alpha\beta}_W
=
\frac{2(n-3)}{(n-2)}(\theta^{\alpha}\wdg *C^\beta-\theta^{\alpha}\wdg *C^\beta)\label{weyl-square-pi}
\ee
for the bivector-valued form $\Pi^{\alpha\beta}_W$. Consequently, the use of variational derivative with respect to connection 1-form in the general formulae (\ref{var-der-general}) immediately yields the vector-valued lagrange multiplier $(n-2)$-form as
\be\label{W-nd-lag-mult-exp}
\lambda^\alpha_W
=
\frac{4(n-3)}{(n-2)}*C^\alpha
\ee
where $C^\alpha=DL^\alpha$ Cotton 2-form and $L^\alpha=R^\alpha-\frac{1}{2(n-1)}R\theta^\alpha$ is the Schouten 1-form in $n$ dimensions introduced above.
The expression (\ref{W-nd-lag-mult-exp}) can be found, by first obtaining
\be
DC^{\alpha\beta}
=
\frac{1}{2}(\theta^\alpha\wdg C^\beta-\theta^\beta\wdg C^\alpha)
\ee
by taking the covariant exterior derivative of (\ref{weyl-def}) and then using the duality relation $*C^{\alpha\beta}=\frac{1}{2}C_{\mu\nu}\wdg *\theta^{\alpha\beta}_{\fant{aa}\mu\nu}$
together with the fact that $D*\theta^{\alpha\beta}_{\fant{aa}\mu\nu}=0$ \cite{phd-heinicke}. Explicitly, in terms of (\ref{weyl-square-pi}),
It is worth stressing  at this point  that $C^\alpha=DL^\alpha$ in the expression (\ref{W-nd-lag-mult-exp}) is four dimensional Cotton 2-form and have
remarkable resemblance to its three dimensional analogue, namely to the corresponding expression for the NMG theory cf. Eqn. (\ref{lag-mult-nmg}) below.
Using the properties of the Weyl 2-form, the energy-momentum forms $*T_W^\mu$ for Weyl 2-form can also be put into the following convenient and simplified form.
Consequently, using all these result, it can be shown that (\ref{weyl-square-eqn-n-dim-form1}) can be rewritten in the form
\be\label{n-dim-weyl-squared-lagrangian}
\frac{4(n-3)}{(n-2)}D*C^\mu
-
\frac{4}{(n-2)}R_\alpha\wdg *C^{\mu\alpha}
+
*T_C^\mu
=0
\ee
where $*T_C^\mu$ is defined  as
\be
*T_C^\mu
\equiv
-(i^\mu C_{\alpha\beta})\wdg*C_{\alpha\beta}+C_{\alpha\beta}\wdg i^\mu *C_{\alpha\beta}.
\ee

Finally, note that for the lagrangian $\frac{1}{m^2}\mathcal{L}_W+R*1$, the field equations can be  written in the form
\be\label{weyl2-eqns-general}
\left(*D*D-\frac{(n-2)}{2(n-3)}m^2\right)R^\alpha
=
\frac{1}{n-3}*(R_\alpha\wdg *C^{\mu\alpha})+(-1)^{s+n}T_C^\mu
\ee
by taking the Hodge dual of the equation. The sign of the last term on the right hand side in (\ref{weyl2-eqns-general}) results from the identity $**=(-1)^{s+p(n-p)}id$ with square of the Hodge dual acting on a $p-$ form in $n$ dimensional space with the metric signature $s$.

\section{Four dimensions}
In four dimensions $n=4$, owing to the fact that the variational derivative of the Gauss-Bonnet lagrangian   vanishes identically, $\delta\mathcal{L}^{(4)}_{GB}=0$, the variational derivatives of the terms $R^2*1$, $R^\alpha\wdg *R_\alpha$ and $\Omega_{\alpha\beta}\wdg *\Omega^{\alpha\beta}$ are not all independent \cite{lanczos-gb}. As a result, the most general form of the QC lagrangians in four dimensions comprises  simply one of the linear combinations $\mathcal{L}_{ik}$, ($i, k=1, 2, 3$) considered above.

It is convenient to consider  the following relation among the QC lagrangians having second order trace
\be\label{generalized-nd-schouten-lagrangian}
\mathcal{L}^{(n)}_{GB}
-
\mathcal{L}^{(n)}_{W}
=
-\frac{2(n-3)}{(n-2)}\left[R^\alpha\wdg *R_\alpha-\frac{n}{4(n-1)}R^2*1\right]
\ee
which holds for $n\geq4$. In four dimensions, this implies that
 \be
 \mathcal{L}_{}^{(4)}
=
R^\alpha\wdg *R_\alpha-\frac{1}{3}R^2*1
=
\mathcal{L}_{12}^{(4)}[a=-\frac{1}{3},b=1]
 \ee
is equivalent to Weyl-squared model. However, it is more convenient to make use of the result of the previous section.
Taking  the identity  satisfied by Weyl 2-form in four dimensions, namely $*T^\mu_C=0$, into account \cite{lovelock}, the metric field equations (\ref{weyl2-eqns-general}) boil down to
\be\label{weyl-kg-form}
\left(*D*D-m^2\right)R^\mu
=
*(R_\alpha\wdg *C^{\mu\alpha})
\ee
where, despite  the fact that curvature tensor in three and four dimensions  have quite distinct properties, it acquires  formal resemblance to the equations of its three dimensional analogue, namely, the NMG theory equations in three dimensions, cf.,  Eqn. (\ref{NMG-aliev-ahmedov-form}) below.  Note that, as has been found above, no QC terms involves in the trace of the equations in four dimensions and as a result, in the present  case, one has $R=0$.  Consequently, (\ref{weyl-kg-form}) also entails Klein-Gordon  type tensorial equation for Ricci 1-form in four dimensions  analogous to the NMG case studied in the following section. However note that it is not easy  to see such a connection using  equivalent lagrangian above.

The linearization of the  field equations  of QC lagrangians supplemented with Einstein-Hilbert term  in the metric perturbations reveals that in four dimensions generically QC theories describe  massless spin-2 modes, together with massive spin-2 modes  in addition to the massive spin-0 mode \cite{stelle}. Although, QC terms renders four dimensional gravity renormalizable, unfortunately
massive spin-2 excitations are ghost and there is a problem with unitarity.  On the other hand, recently proposed Critical gravity model is obtained by the choice of constants $a, b$ in $\mathcal{L}_{12}^{(4)}$ as in the case (1) studied above together with  addition of a particular cosmological term $\lambda*1$.
The choice of $\frac{1}{3}a+b=0$ removes spin-0 mode whereas the cosmological constant with $b+\frac{1}{2\Lambda}=0$ effects  the massive spin-2 modes that they become massless \cite{lu-pope}.

The linearization of field equations (\ref{weyl-kg-form})  around a vacuum solution in terms of metric perturbations immediately yield
an equation with a fourth order linear differential operator which is of the form of a product of two second order differential operators.
In addition, in three dimensions for NMG theory, the second order operator in (\ref{weyl-kg-form}) can also be effectively  \emph{factorized}.
The construction of Critical gravity in four  dimensions also makes the models based on QC lagrangians  studied the cases (2)-(3) more appealing for further study in $n>4$ dimensions towards extension of critical gravity to higher dimensions \cite{sisman}.

\section{Three dimensions}

Three dimensions can be considered as exceptional regarding the field equations for the gravitational theories
that have second order trace. As is well-known, the Einstein-Hilbert action $R*1$ does not have to a propagating degree of freedom and does not lead to a dynamical theory of gravity.  On the other  hand, there are two higher order theories which have desirable properties regarding the particle spectra and perturbative quantum  theory based on these lagrangians.

In this section, the classical aspects of Topologically massive and New massive gravity theories, which are third  and fourth order in the metric components respectively, will be studied in the orthonormal coframe formulation provided above.
It will explicitly be shown that the properties of the NMG field equations as a QC lagrangian are peculiar to three dimensions and in that the field equations NMG field equations are   related to those of TMG theory in a special way under certain conditions.

The QC part of the lagrangian  corresponds to  the case (1) with $\mathcal{L}^{(3)}_{12}[a=-\frac{3}{8}, b=1]$   and customarily denoted by $\mathcal{L}_K$ in three dimensions. In a more common form,   NMG lagrangian  explicitly reads
\be
\mathcal{L}_{NMG}
=
\sigma\mathcal{L}_{EH}
+
\frac{1}{m^2}\mathcal{L}_{K}
=
\sigma \Omega_{\alpha\beta}\wdg*\theta^{\alpha\beta}
+
\frac{1}{m^2}(R^\alpha\wdg *R_\alpha-\frac{3}{8}R^2*1).
\ee
The sign of the Einstein-Hilbert relative to the QC terms has crucial effect on the particle spectrum of the NMG lagrangian, thus
the sign will be encoded in the constant $\sigma$.
As before, the NMG field equations can be written down in the form
\be\label{NMG-eqns-original-form}
-2m^2\sigma*G^\alpha
+
*T^\alpha_{K}
+
D\lambda^\alpha_{K}
=0
\ee
where the auxiliary 2-form $X^{\alpha\beta}_K$, that is to be used in $*T^\alpha_{K}$, takes the following equivalent forms
\begin{eqnarray}
X^{\alpha\beta}_{K}
&=&
\theta^\alpha\wdg (R^\beta-\frac{3}{8}R\theta^{\beta})-\theta^\beta\wdg (R^\alpha-\frac{3}{8}R\theta^{\alpha})
\nonumber\\
&=&
\Omega^{\alpha\beta}-\frac{1}{4}R\theta^{\alpha\beta}
\label{X-tensor-nmg}
\end{eqnarray}
in this case and  using the general formulae (\ref{general-expression-lag-mult}) with
$
\Pi^{\alpha\beta}_{K}
=
D*X^{\alpha\beta}_{K}
$
leads to
\be\label{lag-mult-nmg}
\lambda^\alpha_{K}
=
2*C^\alpha
\ee
where the result is a special case of  the general expression (\ref{lag-mult12-form2}).
The remarkable fact about (\ref{lag-mult-nmg}) is that all the fourth order terms are absorbed into the single expression involving the Hodge dual of Cotton 2-form.  Since it involves the Hodge dual of the Cotton 2-form MNG equations  are parity preserving unlike those of TMG. In contrast to  the field equations of the QC lagrangian (\ref{schouten-squared-lagrangian-form2}) studied in the case (1) in $n\geq4$ dimensions, the fourth order lagrange multiplier term turns out to have the same form  in three dimensions as well. In particular, recall that the simplified form of the lagrange multiplier term in $n\geq4$ dimensions has been obtained by making use of the properties of the Weyl 2-form which in fact defined for dimensions $n\geq4$. The result (\ref{lag-mult-nmg}) for the Lagrange multiplier forms for the NMG theory \cite{baykal-delice} also implies that the NMG field equations can be related to TMG equations  both of them involve Cotton 2-form. Moreover, the Schouten 1-form can be regarded as a kind of potential because of the relation $C^\alpha=DL^\alpha$.

Note that in both  TMG and NMG field equations, higher order terms   involve Cotton 2-form, or equivalently covariant exterior derivatives of the Schouten 1-form. Moreover,  as a consequence of the trace-free property of the Cotton tensor, cosmologically extended TMG field equations has the property  that the trace of the metric field equations are second order as in the case of  NMG. As will explicitly be shown below, such similarities implies closer connection between the mathematical  structures of the two theories.

In three dimensions, it is well-known that there is no Weyl  2-form and the conformal properties of three dimensional metrics are encoded in Cotton 2-form, see, for example, \cite{benn-tucker}. Consequently, it is evident from the NMG equations (\ref{NMG-eqns-original-form}) that for conformally flat solutions, the order of the  NMG equations are in fact
second as a result of the fact that  $C^\mu=0$ for conformally flat solutions.

The NMG field equations written in the from (\ref{NMG-eqns-original-form})  further implies that it is possible to relate them to TMG  field equations when the field equations for two theories are written in terms of Schouten 1-form.
In order to provide such a connection between NMG and TMG, it is first convenient to recall that
TMG field equations  follow from gravitational Chern-Simons action supplemented with Einstein-Hilbert action \cite{djt,cotton}
\be\label{tmg-lag}
\mathcal{L}_{TMG}
=
\sigma R*1
+
\frac{1}{2\mu}(\omega^{\alpha}_{\fant{a}\beta}\wdg d\omega^{\beta}_{\fant{a}\alpha}
+
\frac{2}{3}\omega^{\alpha}_{\fant{a}\beta}\wdg \omega^{\beta}_{\fant{a}\mu} \wdg \omega^{\mu}_{\fant{a}\alpha})
\ee
where a multiplicative constant  $\sigma$ is introduced for convenience.
In terms of differential forms,  the vacuum TMG field equation that follows from this lagrangian  reads \cite{cotton}
\be\label{TMG-original-form}
\sigma*G^\alpha+\frac{1}{\mu}C^\alpha=0
\ee
where $\Pi^{\alpha\beta}=\frac{1}{\mu}\Omega^{\beta\alpha}$  and $\lambda^\alpha=-\frac{2}{\mu}L^\alpha$ for the Chern-Simons part in (\ref{tmg-lag}). The  trace of the vacuum TMG field equations imply that $R=0$ as a result of the fact that the Cotton tensor has vanishing trace $i_\alpha C^\alpha=0$. Now, it will be convenient to rewrite  (\ref{TMG-original-form}) in terms of Schouten 1-form as
\be\label{tmg-eqn-form2}
(*D-\sigma\mu) L^\alpha=0
\ee
by taking the Hodge dual of (\ref{TMG-original-form}). Then by applying the operator $*D$ to  (\ref{tmg-eqn-form2}) and using (\ref{tmg-eqn-form2}) in the resulting expression, one finds
\be\label{TMG-squared}
(*D*D-\sigma^2\mu^2) L^\alpha
=
0
\ee
which are fourth order  in the metric components relative to a coordinate basis. Note also that for $R=0$ it is possible to replace  $L^\alpha$ with $R^\alpha$
in (\ref{TMG-squared}). By linearizing the full nonlinear Eqn.  (\ref{TMG-squared}) by the replacement of $L^\alpha$ with the corresponding linearized Ricci tensor and at the same time by the replacement of the sequence of the covariant exterior derivatives and Hodge duals $*D*D$ with three dimensional Laplacian  of the Minkowski background, one finds that the linearized curvature (in the traceless transverse gauge) represents an excitation spin-2 particle with mass $\mu$ \cite{djt}.

As will explicitly be shown below, the form of the field equations, in fact, suggests that NMG can also be cast into a form analogous to (\ref{TMG-squared}).
In passing note that inclusion of a cosmological constant results in a constant scalar curvature and consequently a covariantly constant source term to (\ref{TMG-squared}). The cosmologically extended TMG equation  then becomes
\be\label{tmg-extended}
(*D*D-\sigma^2\mu^2)L^\alpha=q^\alpha
\ee
where $q^\alpha\equiv\frac{1}{2}\sigma\Lambda_1\mu^2\theta^\alpha$ is covariantly constant vector-valued 1-form: $Dq^\alpha=0$.
In the form given above, the highest derivative term in the \emph{squared} TMG equations contain the term $D*C^\alpha$ which is also present in the  NMG equations (\ref{NMG-eqns-original-form}).

Now, one can  work from NMG side towards (\ref{TMG-squared}) or (\ref{tmg-extended}).
By using the definition $C^\alpha=DL^\alpha$ and also expressing the Einstein forms in terms of Schouten 1-forms and scalar  curvature, NMG field equations (\ref{NMG-eqns-original-form}) can be rewritten in the form
\be
(*D*D+\sigma m^2)L^\alpha
=
\frac{1}{2}\sigma m^2 R\theta^\alpha
+
\frac{1}{2}T_{K}^\alpha
\ee
The source term on the right hand side can further be simplified by using the trace of the field equations. Explicitly, by wedging (\ref{NMG-eqns-original-form}) from the left by the basis coframe form $\theta_\mu$ one finds
\be
2m^2\sigma R*1+\mathcal{L}_K=0.
\ee
Consequently, the NMG field equations take the form
\be\label{NMG-aliev-ahmedov-form}
(*D*D+\sigma m^2)L^\alpha=\frac{1}{2}(T^\alpha_{K}-\frac{1}{2}T_{K}\theta^\alpha)
\ee
where $T_{K}\equiv T^\mu_{K\mu}$ is the trace.
It is worth emphasizing  that, up to this point, neither any  simplifying assumption nor any approximation has been introduced into the derivation of  NMG equation (\ref{NMG-aliev-ahmedov-form}) and it holds in  full generality. The formal  similarity between  (\ref{NMG-aliev-ahmedov-form}) and (\ref{TMG-squared})  becomes even more refined in the cases where
\begin{itemize}
\item[(i)]The source 1-form in (\ref{NMG-aliev-ahmedov-form}) vanishes  identically,
\item[(ii)]It is proportional to Schouten 1-form,
\be\label{nmg-algebraic-constraint}
\frac{1}{2}(T^\alpha_{K}-\frac{1}{2}T_{K}\theta^\alpha)
=
\lambda L^\alpha.
\ee
\end{itemize}
In the latter case, the NMG field equations take the form
\be\label{nmg-last-form}
(*D*D-m'^2) L^\alpha=0.
\ee
where the shifted mass has then been defined  by $-m'^2\equiv\sigma m^2-\lambda$.
Thus, as a result, if a  solution of NMG  satisfies the algebraic constraint (\ref{nmg-algebraic-constraint}), also satisfies the Klein-Gordon type Equation (\ref{nmg-last-form}) and therefore  it satisfies  the ``square" of the TMG equations provided that the mass parameters of the  theories are  simply related by $\mu=m'$.

The foregoing analysis can easily be extended to the case of TMG and/or NMG extended by  respective cosmological constants $\Lambda_1$ and $\Lambda_2$.
For instance, for the latter,  the NMG field equations (\ref{NMG-aliev-ahmedov-form}) extended by  cosmological constant  (via the additional term $-2\Lambda_2*1$ in the NMG action) take the form
\be\label{nmg-extended}
(*D*D+\sigma m^2)L^\alpha=\frac{1}{2}(T^\alpha_{K}-\frac{1}{2}T_{K}\theta^\alpha+\Lambda_2m^2\theta^\alpha).
\ee

An immediate consequence of Eqns. (\ref{nmg-last-form}) and/or (\ref{tmg-extended})-(\ref{nmg-extended})  is that with the algebraic constraints specified above satisfied, the TMG and NMG theories admit  common set of solutions.
Consequently, with the algebraic relation between the mass parameters of the two theories and the condition  relating  expressed in terms Schouten 1-form and QC energy momentum 1-form $T^\mu_K$,  TMG field equations are in fact can be considered as the exact ``square-root" of the NMG field equations. This  correspondence has previously been introduced in \cite{ahmedov-aliev} in a more general context and used to find the solutions of NMG field equations using the field equations of TMG in a series of papers \cite{ahmedov-aliev}. The presentation given here in terms of exterior calculus of forms allows one to identify the Dirac type operator $\slashed{D}$ introduced in  \cite{ahmedov-aliev} simply as $\slashed{D}=*D$ in terms of  geometrical operators, namely, the covariant exterior  derivative in composition with the Hodge dual relative to an orthonormal coframe.

Now it is a convenient point to comment briefly  on the equations for General massive gravity (GMG) in the  framework above.
The  GMG theory follows from the lagrangian which is a combination of  those of TMG and NMG  theories and thus  has two distinct mass parameters \cite{NMG-bth}. TMG and NMG theories can be recovered by taking appropriate limits of the two mass parameters of GMG. It is convenient to consider directly the field equations for GMG which are of the form
\be\label{gmg-eqn-form1}
\sigma *G^\alpha+\frac{1}{\mu}C^\alpha+\frac{1}{m^2}\left(D*C^\alpha+*T^\alpha_{K}\right)=0.
\ee
By introducing  the mass parameters $m_{\mp}$ for GMG by redefinitions
$m^2\equiv m_+m_-$ and $\frac{m^2}{\mu}\equiv m_++m_-$,
it is easy to show that the GMG equations take the form
\be\label{gmg-eqn-form2}
[*D*D-(m_++m_-)*D+m_+m_-]L^\alpha
=
T^\alpha_K
\ee
in terms of the new mass parameters with the QC source term $T^\alpha_K$ is defined as  the same as that of the NMG theory. Note that
(\ref{gmg-eqn-form2}) is not a linearized equation in metric perturbation and holds in full generality.  Thus, it is not possible to write the differential operator on the left hand side   in a factorized form. On the other hand, the equation provides a convenient expression with regard to  linear
approximation in the metric perturbation around a flat background. Explicitly, the leading term on the  left hand side of (\ref{gmg-eqn-form2})
is of the form
\begin{eqnarray}
[*D*D-(m_++m_-)*D+m_+m_-]L^\alpha
&=&
[*d*d-(m_++m_-)*d+m_+m_-]L^\alpha
+\ldots
\nonumber\\
&\approx&
(*d-m_+)(*d-m_-)L^\alpha_{lin.}+\ldots
\end{eqnarray}
where the Hodge duals on the right hand side on the second line are assumed  to be approximated by the Hodge duals of the flat
background. The factorized differential operators of the flat background act on linearized Schouten 1-form $L^\alpha_{lin.}.$ As the GMG calculations elegantly illustrate,  the mathematical formalism presented allows to obtain the  linear terms in a factorized form and in fact these considerations  apply to   the other QC field equations above.

Finally, recall that a constant curvature spacetime, with curvature 2-form,
\be\label{symmetric-curvature-2form}
\Omega^{\alpha\beta}=\frac{-2\Lambda}{(n-1)(n-2)}\theta^{\alpha\beta},
\ee
satisfies the Einstein field equations $*G^\alpha=\Lambda*\theta^\alpha$. For the  general lagrangian (\ref{n>4-qc-lag}) supplemented with
Einstein-Hilbert action and a cosmological constant $\Lambda_0$, i.e., for the lagrangian of the form
\be\label{extended-lag}
\mathcal{L}^{(n)}
=
\frac{1}{\kappa}(\Omega_{\alpha\beta}\wdg*\theta^{\alpha\beta}-2\Lambda_0*1)+\mathcal{L}^{(n)}_t,
\ee
the Einstein metric   of (\ref{symmetric-curvature-2form}) produces
\be
*T^{\alpha}_t
=
\Lambda^2\frac{4(n-4)}{(n-2)^2}\left(an+b+\frac{c}{(n-1)}\right)*\theta^\alpha
\ee
which can be regarded as a kind of cosmological term. In addition, one has $D\lambda^\alpha_t=0$ identically. Eventually, with the maximally symmetric metrics,
the vacuum field equations for (\ref{extended-lag}) reduce to
\be
*G^\alpha+\left[\Lambda_0-\Lambda^2\frac{2(n-4)}{\kappa(n-2)^2}\left(an+b+\frac{c}{(n-1)}\right)\right]*\theta^\alpha
=0.
\ee
Thus, if the cosmological constant satisfies the equation
\be\label{quadratic-cosmo}
\frac{1}{2\kappa}(\Lambda-\Lambda_0)+\Lambda^2\frac{(n-4)}{(n-2)^2}\left(an+b+\frac{c}{(n-1)}\right)=0
\ee
then, (\ref{symmetric-curvature-2form})  also solves the vacuum equations of the QC model (\ref{extended-lag}) \cite{deser-tekin}.
However, now there are two distinct vacuum solutions corresponding to the roots of Eqn. (\ref{quadratic-cosmo}). Thus, generic
QC field equations can be linearized around  the maximally symmetric vacua as well as flat vacuum background.

\section{Concluding remarks}
In the above study of QC gravity equations,  a concise form of the general QC gravity equations (\ref{4+d-field-eqn}) relative to orhonormal coframe are derived. These  provide a different point of view and  further insight into the QC field equations. They can be cast into various other forms that may be useful for some other purposes as well. In this regard, note that (\ref{weyl-def}) can  be rewritten in the form
\be
\Omega^{\alpha\beta}
=
\sum_{k=1}^{3}W_k^{\alpha\beta}
\ee
which shows explicitly the irreducible  decomposition of the curvature 2-form \cite{hehl}.
$W_1^{\alpha\beta}$ is traceless fourth rank part $W_1^{\alpha\beta}\equiv C^{\alpha\beta}$. $W_2^{\alpha\beta}$ is traceless Ricci part
$W_2^{\alpha\beta}=\frac{1}{(n-2)}(\theta^{\alpha}\wdg S^\beta-\theta^{\beta}\wdg S^\alpha)$ where $S^\alpha\equiv R^\alpha-\frac{1}{n}R\theta^\alpha$
is the traceless Ricci 1-form and finally, $W_3^{\alpha\beta}\equiv \frac{1}{n(n-1)}R\theta^{\alpha\beta}$ is the trace part.
Then, using this decomposition of the curvature 2-form, the second equality  in (\ref{weyl-square-eqn-n-dim-form1}) can be written in the form
\be
\Omega^{\alpha\beta}\wdg*\Omega_{\alpha\beta}
=
\sum_{k=1}^{3}W_k^{\alpha\beta}\wdg *{W_k}_{\alpha\beta}.
\ee
Moreover, the irreducible parts are orthogonal in the sense that
$
\Omega^{\alpha\beta}\wdg*{W_k}_{\alpha\beta}
=
{W_k}^{\alpha\beta}\wdg *{W_k}_{\alpha\beta}
$
and ${W_k}^{\alpha\beta}\wdg *{W_j}_{\alpha\beta}\propto \delta_{jk}$. Now, it is easy to formulate the QC models by employing the lagrangian densities
of the form ${W_k}^{\alpha\beta}\wdg *{W_j}_{\alpha\beta}$. The corresponding bivector valued 2-form then takes the form
 $X_k^{\alpha\beta}=2{W_k}^{\alpha\beta}$.  Therefore, in terms of these, the field equations (\ref{4+d-field-eqn}) directly expresses the field equations in terms of the contractions  of the irreducible parts of the curvature. Consequently, (\ref{4+d-field-eqn}) can be put in a convenient to form to write the general QC field equations relative to the null coframe   of Newman-Penrose   in terms of spinor components \cite{NP,nutku-aliev}.

Apparently, the construction of  QC lagrangians having second order trace equations in terms  of the bilinear combinations considered above
requires the coupling constants (free parameters) that depend crucially on the number of spacetime dimensions. As a consequence, a QC lagrangian density having this property for a given  spacetime dimensions does not generalize to other dimensions. Regarding the trace property, each spacetime dimension has to be considered separately. In four dimensions, for example, the most general QC gravitational lagrangian that has second order trace is of the form (\ref{schouten-squared-lagrangian-form2}) with particular values of the coupling constants $a=-\frac{1}{3}b$.  However, for $n>4$, the most general QC lagrangian contains all the three terms considered in the cases (1)-(3) above that cannot be eliminated by adding a topological  term as in  four dimensions. Therefore, there are three independent linear combinations of QC lagrangians each of which  has the desired property of second order trace and consequently, the most general  QC lagrangian contains three free parameters. In dimensions $n>4$,  the QC lagrangians derived in cases (2) and (3) above deserve further scrutiny with regard to, for example, their particle spectra and classical solutions as well.

One of the original results of the paper is that,  the orthonormal coframe formulation of the field equations allows the lagrangians of the  form (\ref{generalized-nd-schouten-lagrangian}) to be cast into Klein-Gordon type tensorial equation of the general form (\ref{NMG-aliev-ahmedov-form}) in any dimensions $n\geq 3$. Three dimensions seem to have peculiar properties that has no analogues in higher dimensions. TMG and NMG field equations are related in a special way, in that, with certain conditions satisfied, the former, which can be written as a Klein-Gordon type equation considered as the square-root of the latter. The direct approach of solving QC gravity equations starting with a suitable metrical ansatze is difficult, however,
the study of coframe formulation of QC field equations may provide insight into the QC field equations. In this regard,
further attention is required to provide concrete examples of class of metrics that satisfy the conditions  either (i) or (ii) which makes the TMG equations square root of the NMG equations.

Another original result is the construction of general QC models having second order trace in $n>4$ dimensions with three free parameters. Consequently,
as can be observed from Eqn. (\ref{original-n-d-lag}), some these QC models involve contraction of Riemann tensor with itself and they have no analogues in $n\leq4$ dimensions.  For these models, the field equations  cannot be written in terms of Schouten tensor as opposed to the well-known QC models involving linear combinations of  Ricci-squared and scalar curvature-squared terms in $n\geq3$ dimensions.

In four dimensions, it may provide further insight for the field equations of the Weyl-squared model  if the field Eqn. (\ref{weyl-kg-form}) can be rewritten in terms of the Lanczos potential for the Weyl 2-form \cite{lanczos-potential}. This, however in the present framework can naturally be achieved by making use of the Lanczos potential formulated relative to  an orthonormal coframe and this requires separate investigation.
The field equations expressed  in terms of the exterior differential forms provide further insight, at the least, into the NMG field equations in relation to TMG theory.  Written down explicitly in terms of differential forms, and  also in terms of Schouten 1-form, the novel description of the NMG field equations \cite{ahmedov-aliev} amounts to taking Hodge dual  of the metric field equations and subsequently regrouping the QC terms conveniently. This result follows from the fact  that relative to orthonormal coframe all the fourth order terms in the field equations can concisely be expressed in terms of Cotton 2-form for the NMG theory. In contrast, the novel description of NMG in terms of traceless Ricci tensor given in \cite{ahmedov-aliev}  constructed by using  identities satisfied by three dimensional curvature tensor.

Finally, it is interesting to note that the second order covariant differential operator $D*D*$ which comes into play by construction of the NMG field equations relative to an orthonormal coframe also shows up in  a quite different  context   of introducing mass to graviton in other gravitational models. In \cite{mielke}, using the first order formalism \cite{hehl}  in three dimensions and in the context topological Riemann-Cartan type theory generalizing TMG, the basis coframe 1-forms are shown to satisfy a Klein-Gordon type equation of the  same form as (\ref{nmg-last-form}) in terms of the differential operator $*D*D\mp D*D*$.

\begin{acknowledgements}
The author would like to thank \"Ozg\"ur Delice for help and support.
The paper is dedicated to U\u gur Feza Baykal.
\end{acknowledgements}


\begin{thebibliography}{99}

\bibitem{weyl}
Weyl, H.:, Gravitation und Elektrizit\"{a}t, Preuss. Akad. Wiss. Berlin, Sitz., 465-480, (1918);
Weyl, H.: Eine neue Erweiterung der Relativit\"{a}tstheorie, Ann. der Phys. {\bf59}  101-133, (1919);
Weyl, H.: Raum-Zeit-Materie, 4th Ed., Springer-Verlag, Berlin (1921).

\bibitem{eddington}
Eddington, A.S.: Relativit\"{a}tstheorie in mathematischer Behandlung,  Berlin: J Springer, (1925).

\bibitem{hj-scmidt-history}
Schmidt, H. -J.:  Fourth order gravity: equations, history, and applications to cosmology. Int. J. Geom. Meth. Mod. Phys. {\bf4}  209-248, (2007); 	 arXiv:gr-qc/0602017v2

\bibitem{zweibach}
Zweibach, B.: Curvature squared terms and string theories. Phys. Lett. {\bf156B}, 315 (1985);
Tseytlin, A.: Ambiguity in the effective action in string theories. Phys. Lett. {\bf156B}, 176, 92 (1986).


\bibitem{odintsov-renor}
Buchbinder, I. L.,  Odintsov,  S. D.  and Shapiro, I. L.:
Effective action in quantum gravity, IOP Publishing, Bristol (1992)

\bibitem{starobinski}
Starobinsky, A.: A new type of isotropic cosmological models without singularity. Phys. Lett. {\bf 91B}, 99 (1980).


\bibitem{kerner}
Kerner, R.: Cosmology without singularity and nonlinear gravitational Lagrangians.  Gen. Relativ. Gravit. {\bf14}, 453 (1982).

\bibitem{jakubiec}
Jakubiec, A. and Kijowski, J.:
 On theories of gravitation with nonlinear Lagrangians. Phys. Rev. D {\bf37}, 1406–1409 (1988);
 Whitt, B.: Fourth Order Gravity as General Relativity Plus Matter. Phys.Lett. {\bf B}145, 176 (1984)

\bibitem{odintsov3}
Shin'ichi, N. and  Odintsov, S. D.:
Modified gravity with negative and positive powers of curvature: Unification of inflation and cosmic acceleration.
Phys. Rev. D 68, 123512 (2003)


\bibitem{djt}
Deser, S., Jackiw R. and Templeton S.: Three-Dimensional Massive Gauge Theories. Phys. Rev. Lett. {\bf48}, 975 (1982);
Deser, S., Jackiw R. and Templeton S.: Topologically massive gauge theories. Ann. Phys., NY {\bf140}, 372–411 (1982); erratum-ibid. {\bf185},  406 (1988)

\bibitem{NMG-bth}
Bergshoeff, E.A., Hohm, O. and Townsend, P.K.: Massive Gravity in Three Dimensions. Phys. Rev. Lett. {\bf102},  201301 (2009); 	arXiv:0901.1766v3 [hep-th]




\bibitem{lu-pope}
L\"{u}, H.and Pope, C.N.:
Critical Gravity in Four Dimensions. Phys. Rev. Lett. {\bf106}, 181302 (2011);
Bergshoeff, E.A., Hohm, O., Rosseel, J. and Townsend, P.K.:
Modes of log gravity. Phys. Rev. D {\bf83}, 104038 (2011)

\bibitem{nakasone-oda}
Nakasone, M. and Oda, I.: 	On Unitarity of Massive Gravity in Three Dimensions. Prog. Theor. Phys. {\bf121}, 1389-1397 (2009); 	arXiv:0902.3531v5 [hep-th]

\bibitem{ray}
Oliva, J. and  Ray S.: Classification of Six Derivative Lagrangians of Gravity and Static Spherically Symmetric Solutions. Phys. Rev. D {\bf82}, 124030  (2010); arXiv:1004.0737v3 [gr-qc]

\bibitem{odintsov1}
Shin'ichi, N. and  Odintsov, S. D.: Introduction To Modified Gravity And Gravitational Alternative For Dark Energy. International Journal of Geometric Methods in Modern Physics. {\bf4} (2007) 115-145.

\bibitem{odintsov2}
Shin'ichi, N. and  Odintsov, S. D.: Unified cosmic history in modified gravity: From F(R) theory to Lorentz non-invariant models.
Physics Reports. {\bf505}, (2011) 59–144.

\bibitem{thirring}
Thirring, W.:  Classical Mathematical Physics: Dynamical Systems and Field Theories. 3rd Ed.  Springer-Verlag New York, Inc.   (2003)

\bibitem{straumann}
Straumann, N.: General Relativity :With Applications to Astrophysics (Theoretical and Mathematical Physics). Springer-Verlag Berlin Heidelberg, (2010)

\bibitem{lovelock}
Lovelock, D.: The Einstein Tensor and Its Generalizations.  J. Math. Phys. {\bf12}, 498  (1971)

\bibitem{baykal-delice}
Baykal, A. and  Delice, \"O.: A Unified Approach to Variational Derivatives of Modified Gravitational Actions. Class. Quant. Grav. {\bf28},
015014 (2011); arXiv:1012.4246v1 [gr-qc]

\bibitem{var-dereli-tucker}
Dereli, T. and  Tucker, R.W.: Variational methods and effective actions in string models. Class. Quant. Grav. {\bf4},  791 (1987)

\bibitem{tekin-deser}
Deser, S. and Tekin, B.: Gravitational Energy in Quadratic-Curvature  Gravities. Pyhs. Rev. Lett. {\bf 89}, 101101-1, (2002)

\bibitem{benn-tucker}
Benn, I.M. and Tucker, R.W.: An Introduction to Spinors and Geometry With Applications in Physics.
Adam and Hilger,  Bristol and Philadelphia (1988)

\bibitem{deser}
Deser S.: Ghost-Free, Finite, Fourth-Order 3D Gravity. Phys. Rev. Lett. {\bf103},   101302 (2009); arXiv:0904.4473v3 [hep-th]


\bibitem{heinicke}
 Heinicke, C.: The Einstein 3-form $G_\alpha$ and its equivalent 1-form $L_\alpha$ in Riemann-Cartan space. Gen. Rel. Grav. {\bf33},  1115-1130 (2001);	 arXiv:gr-qc/0012037v1

\bibitem{phd-heinicke}
Heinicke, C.: Exact solutions in Einstein's theory and beyond. Ph.D. Thesis, University of Cologne, (2005)

\bibitem{lanczos-gb}
Lanczos, C.: A Remarkable Property of the Riemann-Christoffel Tensor in Four Dimensions. Ann. Math., NY {\bf39}, 842 (1938)

\bibitem{cotton}
Garcia, A.,  Hehl F.W.,  Heinicke C. and  Macias A.: The Cotton tensor in Riemannian spacetimes. Class. Quant. Grav. {\bf21}  1099-1118, (2004); 	 arXiv:gr-qc/0309008v2

\bibitem{bach}
Bach, R.:  Zur Weylschen Relativit\"{a}tstheorie und der Weylschen Erweiterung
des Kr\"{u}mmungsbegriffs. Math. Zeitschr. {\bf9},  110-135 (1921)

\bibitem{fiedler-schimming}
Fiedler, B. and  Schimming R.: Exact Solutions Of The Bach Field Equations Of General Relativity. Reports on Mathematical Physics, {\bf17},  15-36 (1980)

\bibitem{dereli-tucker-weyl}
Dereli, T. and Tucker, R.W.:  A note on generalisation of Weyl's theory of gravitation. J. Phys. A:  Math. Gen. {\bf15},  L7-L11 (1982)

\bibitem{sisman}
Deser S., Liu, H., Lu H.,  Pope C.N., \c{S}i\c{s}man, T.\c{C}., and  Tekin, B.:
Critical points of D-dimensional extended gravities. Phys. Rev. D {\bf83}, 061502 (2011)
L\"u H., Pang Y. and Pope C.N.:
Conformal gravity and extensions of critical gravity. Phys. Rev. D {\bf84}, 064001 (2011)



\bibitem{kopczynsky}
 Candelas, P.,  Horowitz, G.T.,  Strominger, A. and  Witten, E.: Vacuum configurations for superstrings. Nucl. Phys. B{\bf258}, 46–74  (1985);
Kopczy\`{n}sky, W.: Variational principles for Gravity and Fluids, Annals of Physics, {\bf 203}, 308-338 (1990)


\bibitem{hehl}
Hehl, F.W.,  McCrea, J.D., Mielke E.W., and Ne'eman Y.:   Metric-affine gauge theory of gravity field equations,
Noether identities, world spinors, and breaking of dilation invariance Phys. Rep. {\bf258}, 1 (1995)


\bibitem{deser-tekin}
Deser, S. and Tekin, B.: Gravitational Energy in Quadratic Curvature Gravities. Phys. Rev. Lett. {\bf89}, 101101 (2002)



\bibitem{ahmedov-aliev}
Ahmedov, H., Aliev, Alikram N.: Type D Solutions of 3D New Massive Gravity. Phys. Rev. D {\bf83}, 084032 (2011)  	arXiv:1103.1086v1 [hep-th];
Ahmedov, H., Aliev, Alikram N.: The General Type N Solution of New Massive Gravity. Phys Lett. B {\bf694}, 143-148 (2010)  arXiv:1008.0303v2 [hep-th];
Ahmedov, H., Aliev, Alikram N.: Exact solutions in 3D New massive  Gravity. Phys. Rev. Lett. {\bf106}, 021301 (2011) ;
Ahmedov, H., Aliev, Alikram N.:, Type D Solutions of 3D New Massive Gravity. Phys. Rev. D {\bf83}, 084032 (2011) arXiv:1103.1086v1 [hep-th]

\bibitem{NP}
Newman, E., and Penrose,  R.:An Approach to Gravitational Radiation by a Method of Spin Coefficients. J. Math. Phys. {\bf3}, 566  (1962)


\bibitem{nutku-aliev}
Aliev, A. N., Nutku, Y,:
Spinor formulation of topologically massive gravity. Class. Quantum Grav. {\bf12}, 2913 (1995)

\bibitem{lanczos-potential}
Lanczos, C.: The splitting of the Riemann tensor, Rev. Mod. Phys. {\bf34}, 379-389 (1962)


\bibitem{mielke}
Mielke, E.W.  and  Baekler, P.: Topological gauge model of gravity with torsion. Phys. Lett. A, {\bf 156}, 399 (1991);
Baekler, P., Mielke E.W., and Hehl F.W.: Dynamical symmetries in topological 3D gravity with torsion. Nuovo Cim., B{\bf107},  91 (1992);
Mielke, E.W. and Rincon Maggiolo, A.A.: Rotating black hole solution in a generalized topological 3-D gravity with torsion.
Phys. Rev. D {\bf68}, 104026 (2003)

\end{thebibliography}
\end{document}